\documentclass[aps,prl,reprint,superscriptaddress,floatfix,amsmath,amssymb,bm]{revtex4-2}
\usepackage{graphicx}
\usepackage{xcolor}
\usepackage{comment} 
\usepackage{physics}
\usepackage{diagbox}
\usepackage[colorlinks,linkcolor=blue,citecolor=blue,urlcolor=blue]{hyperref}

\newcommand{\iu}{\mathrm{i}}
\newcommand{\eu}{\mathrm{e}}
\usepackage{siunitx}
\usepackage[normalem]{ulem}

\begin{document}
\title{Carrollian Flat Bands: Quantum Geometry and a Hierarchy of Immobility}

\author{Satoshi Iso}
\affiliation{RIKEN Center for Interdisciplinary Theoretical and Mathematical Sciences (iTHEMS), RIKEN, Wako, Saitama 351-0198, Japan}
\affiliation{The Graduate University for Advanced Studies (SOKENDAI)}
\affiliation{Institute of Particle and Nuclear Studies, High Energy Accelerator Research Organization (KEK)}
\author{Puttarak Jai-akson}
\affiliation{RIKEN Center for Interdisciplinary Theoretical and Mathematical Sciences (iTHEMS), RIKEN, Wako, Saitama 351-0198, Japan}
\affiliation{Department of Physics, Kyoto University, Kitashirakawa-Oiwakecho, Kyoto 606-8502, Japan}
\author{Tomoki Ozawa}
\affiliation{Advanced Institute for Materials Research (WPI-AIMR), Tohoku University, Sendai 980-8577, Japan}
\affiliation{RIKEN Center for Interdisciplinary Theoretical and Mathematical Sciences (iTHEMS), RIKEN, Wako, Saitama 351-0198, Japan}

\date{\today}

\begin{abstract}
Carrollian symmetry, the symmetry of the limit in which the speed of light vanishes, is the counterpart of the non-relativistic limit in high-energy physics. Since nothing can move in this limit, Carrollian systems are naturally compared with flat bands, in which the group velocity vanishes. We make this analogy precise and show that it is more restrictive than flatness. Carrollian symmetry requires the boost generators, which implement position-dependent time translations, to commute, which on a flat band is equivalent to a vanishing Berry curvature. The stronger ultra-local symmetry, under which all position-dependent time translations commute, is equivalent to a vanishing minimal quantum metric and singles out bands spanned by compact localized states with disjoint supports. These conditions define a hierarchy of immobility. In a generic flat band a wave packet moves under a force through the anomalous velocity; in a Carrollian flat band a uniform force produces no center-of-mass motion within the band, but interactions can still displace particles; in an ultra-local Carrollian flat band nothing moves at all. In the last case the interacting many-body problem reduces exactly to a classical lattice gas even on a connected lattice, Baxter's hard-hexagon model for a triangular-lattice example, with corrections away from this point that are proportional to the minimal quantum metric of the band.
\end{abstract}

\maketitle

Flat bands, which are bands whose dispersion is momentum independent, have been a source of various exotic phenomena in condensed matter physics, because in such bands the kinetic energy plays no role, and the many-body physics is determined by the inter-particle interactions. The emergence of itinerant ferromagnetism in flat bands~\cite{Mielke1991a,Mielke1991b,Tasaki1992,MielkeTasaki1993,Tasaki1998} and the fractional quantum Hall effects~\cite{Tsui1982,Laughlin1983,Neupert2011,Sheng2011,Regnault2011,Parameswaran2013,Bergholtz2013} are representative phenomena where interaction effects on flat bands lead to curious many-body phases. The recent discovery of nearly flat bands in twisted bilayer graphene and other moir\'e systems~\cite{Bistritzer2011,Cao2018a,Cao2018b,Andrei2020} has led to renewed interest in the physics of flat bands.

Independently of the study of flat bands, the concept of Carrollian symmetry has been proposed and developed in high-energy physics~\cite{LevyLeblond1965,SenGupta1966,Henneaux1979,Duval2014carroll,Bergshoeff2014,deBoer2022}. The Carrollian symmetry is the symmetry under the Carroll boost $ t \to t - \boldsymbol{\beta}\cdot\mathbf{x}$, which arises in the limit where the speed of light $c$ goes to zero, $c \to 0$.
If one thinks of the non-relativistic limit as the $c \to \infty$ limit, the Carrollian limit is its exact opposite.
In the limit $c \to 0$ a free massive particle cannot move~\cite{Bergshoeff2014,FigueroaOFarrill2023}. Because of this ultra-locality, Carrollian theories typically enjoy an infinite-dimensional extension of the boost symmetry, the supertranslations $t \to t - f(\mathbf{x})$ with an arbitrary function $f(\mathbf{x})$~\cite{Henneaux1979,Duval2014bms}. Recently, a few works have appeared relating the physics of the Carrollian limit to condensed matter physics, especially to flat-band systems~\cite{Bagchi:2023:JHEP,Ara:2024:SciPost,Banerjee2026}, since in both cases particles have zero (group) velocity. In particular, Ref.~\cite{Ara:2024:SciPost} showed in one dimension that supertranslation invariance of the Hamiltonian leads to the existence of compact localized states, and hence to flat bands.

In this paper we ask which flat bands can really be called Carrollian. Having a flat dispersion alone does not imply that the system realizes the Carrollian symmetry, in the sense that the Carroll group is represented on the flat band, with generators of the boosts that are conserved \emph{and} close into the Carroll algebra. We find that this holds if and only if the Berry curvature vanishes throughout the Brillouin zone. Furthermore, its infinite-dimensional extension to supertranslations $t\to t-f(\mathbf x)$, which characterizes ultra-local Carrollian theories, is realized if and only if the minimal quantum metric vanishes. Such bands are spanned by compact localized states with disjoint supports. 
These two conditions define a hierarchy of immobility. In a generic flat band the group velocity vanishes but a wave packet still moves under a uniform force through the anomalous velocity. In Carrollian flat bands this anomalous velocity is absent, so that within the band the center of mass is unaffected by a uniform force, but interactions can set it in motion. In ultra-local Carrollian flat bands even interactions cannot make the particles move, and the interacting many-body problem is exactly a classical lattice gas, e.g. Baxter's hard-hexagon model~\cite{Baxter1980,Baxter1982} for a triangular lattice, with corrections away from the ultra-local point proportional to the minimal quantum metric. Carrollian symmetry underlies a range of phenomena in high-energy physics, from flat-space holography and the physics of black-hole horizons~\cite{Duval2014bms,Bagchi2026} to Carrollian cosmology~\cite{deBoer2022} and the Carroll/fracton correspondence~\cite{FigueroaOFarrill2023}. Our work opens a way to find their analogues in flat-band systems realizable in condensed-matter and cold-atom platforms.

\section{Carrollian flat band}
A system is Carrollian if the Carroll group is represented on it. 
There are boost generators $B_i$ which commute with the Hamiltonian and with each other. 
They transform under a lattice translation $T_{\mathbf a}$ by a lattice vector ${\mathbf a}$ as $T_{\mathbf a}^\dagger B_i T_{\mathbf a}=B_i+a_iH$. A brief review of the Carroll algebra and its infinite-dimensional extension is provided in the End Matter.
We now construct the boost generators of a flat band and ask when these relations hold.
Under the Carroll boost $t \to t - \boldsymbol{\beta}\cdot\mathbf{x}$, the wavefunction transforms as $U(\boldsymbol\beta)\psi(\mathbf{x},t) = \psi(\mathbf{x},t + \boldsymbol\beta \cdot \mathbf{x}) \approx \psi(\mathbf{x},t) + \boldsymbol\beta \cdot \mathbf{x} \partial_t \psi(\mathbf{x},t) = \left( 1 - \iu \boldsymbol\beta \cdot \mathbf{x} H\right)\psi(\mathbf{x},t)$. Here, $t + \boldsymbol\beta \cdot \mathbf{x}$ is the original time written in terms of the transformed time, and we used the Schr\"odinger equation $\iu\partial_t \psi = H\psi$ in the last step.
Thus, the boost generator $B_i$, defined by $U(\boldsymbol\beta) \approx (1 - i\boldsymbol\beta \cdot \mathbf{B})$ should be $B_i = x_i H$ or more precisely its Hermitian form $B_i = (x_i H + H x_i)/2$.

Let us now consider an isolated flat band with Bloch states $|\psi_{\mathbf k}\rangle$ and projector $P=\sum_{\mathbf k}|\psi_{\mathbf k}\rangle\langle\psi_{\mathbf k}|$, so that $HP=PH=E_0P$ with $E_0\neq0$ the flat-band energy; the projected Hamiltonian is $\bar H\equiv PHP=E_0P$.
The boost generator $B_i$ restricted to the flat band is
\begin{align}
    \bar{B}_i &\equiv P B_i P = E_0 P x_i P \notag \\
    &=
    E_0 \sum_\mathbf{k} |\psi_\mathbf{k}\rangle \left( \iu \nabla_\mathbf{k} + \boldsymbol{\mathcal{A}}(\mathbf{k}) \right)_i \langle \psi_\mathbf{k}| \label{eq:barbi}
\end{align}
where $\boldsymbol{\mathcal{A}}(\mathbf{k}) \equiv \iu \langle u_\mathbf{k}|\nabla_\mathbf{k} u_\mathbf{k}\rangle$ is the Berry connection of the band~\cite{Berry1984,Xiao2010} with $|u_\mathbf{k}\rangle \equiv \eu^{-\iu\mathbf{k}\cdot \mathbf{x}}|\psi_\mathbf{k}\rangle$ being the cell-periodic part of the Bloch state $|\psi_\mathbf{k}\rangle$. Note that the derivative $\nabla_\mathbf{k}$ in \eqref{eq:barbi} acts on everything to its right.
We note that Eq.~\eqref{eq:barbi} shows that the Carroll boost of a flat band is the band energy times the projected dipole moment $P x_i P$, which also connects flat bands to the Carroll/fracton correspondence~\cite{FigueroaOFarrill2023}, in which the Carroll boosts play the role of the
conserved dipole moment of a fracton system.
The nontrivial condition for these projected boost generators to realize the Carroll algebra is that they commute with each other. Using Eq.~\eqref{eq:barbi}, this implies
\begin{align}
    \left[ \bar{B}_i,\bar{B}_j \right]
    =
    E_0^2 \sum_\mathbf{k}|\psi_\mathbf{k}\rangle \iu\Omega_{ij}(\mathbf{k}) \langle \psi_\mathbf{k}|
    =
    0,
    \label{eq:boostcomm}
\end{align}
where $\Omega_{ij}(\mathbf{k}) \equiv \partial_{k_i}\mathcal{A}_j (\mathbf{k}) - \partial_{k_j}\mathcal{A}_i (\mathbf{k})$ is the Berry curvature of the band. 
The Carroll algebra requires the boost generators to commute. Equation~\eqref{eq:boostcomm} therefore shows that the projected boosts satisfy this condition if and only if $\Omega_{ij}(\mathbf{k})=0$ throughout the Brillouin zone.
We call flat bands satisfying this condition \textit{Carrollian flat bands}.
We note that the boost generator $\bar{B}_i$ commutes with $\bar{H}$ and also with $H$.
The remaining relation of the Carroll algebra, $T_{\mathbf a}^\dagger\bar B_iT_{\mathbf a}=\bar B_i+a_i\bar{H}$, holds automatically for any flat band, since $T_{\mathbf a}$ commutes with $P$ and $T_{\mathbf a}^\dagger\mathbf xT_{\mathbf a}=\mathbf x+\mathbf a$. This is the lattice form of the commutation relation $[B_i,P_j]=\iu \delta_{ij}H$ of the continuum Carroll algebra, with $P_j$ the generator of translations in the $j$-th direction.
For $\Omega_{ij}\neq0$ the projected boosts instead form an extension of the Carroll algebra, 
central extension for a uniform curvature \cite{Marsot2022}  and non-central otherwise (see Supplemental Material).

\section{Ultra-local Carrollian flat band}

We now ask when the flat band is invariant under the more general supertranslation
$t\to t-f(\mathbf x)$ with an arbitrary function $f(\mathbf{x})$. In the language of Carrollian physics, such systems are called \textit{ultra-local}.
Repeating the argument above, we can take the generator of the supertranslation to be
$Q_f = f(\mathbf{x})H$, or, again, more precisely its Hermitian form $Q_f = \left( f(\mathbf{x})H + H f(\mathbf{x}) \right)/2$. 
Projecting onto the flat band,
\begin{equation}
  \bar Q_f\equiv P Q_f P = E_0\,P f(\mathbf x) P . \label{qf}
\end{equation}
We require any two projected generators to commute: $[\bar Q_f, \bar Q_g] = 0$. To understand the condition more explicitly, we expand $f(\mathbf{x})$ in plane waves as $f(\mathbf x)=\sum_{\mathbf q}\tilde f(\mathbf q)\eu^{\iu\mathbf q\cdot\mathbf x}$. The condition $[\bar Q_f, \bar Q_g] = 0$ is now equivalent to the condition that the projected density operators
\begin{equation}
  \bar n_{\mathbf q}\equiv P \eu^{\iu\mathbf q\cdot\mathbf x}P
  =\sum_{\mathbf k}|\psi_{\mathbf k+\mathbf q}\rangle\,\Lambda_{\mathbf{k}+\mathbf{q},\mathbf{k}}\,\langle\psi_{\mathbf k}|,
  \label{eq:nbar}
\end{equation}
commute, where $\Lambda_{\mathbf{k}+\mathbf{q},\mathbf{k}} \equiv \langle u_{\mathbf k+\mathbf q}|u_{\mathbf k}\rangle$ is the form factor of the flat band~\cite{Girvin1986,Parameswaran2012,Roy2014}. The commutator is
\begin{align}
  &0 = [\bar n_{\mathbf q},\bar n_{\mathbf q'}]=  
  \sum_{\mathbf k}|\psi_{\mathbf k+\mathbf q+\mathbf q'}\rangle
  \label{eq:densityalgebra}\\
  &\left[\Lambda_{\mathbf{k}+\mathbf{q}+\mathbf{q}^\prime,\mathbf{k}+\mathbf{q}^\prime}\Lambda_{\mathbf{k}+\mathbf{q}^\prime,\mathbf{k}}
  -\Lambda_{\mathbf{k}+\mathbf{q}+\mathbf{q}^\prime,\mathbf{k}+\mathbf{q}}\Lambda_{\mathbf{k}+\mathbf{q},\mathbf{k}}\right]
  \langle\psi_{\mathbf k}| \notag
\end{align}
for any $\mathbf{q}$ and $\mathbf{q}^\prime$.
We call flat bands where this condition is satisfied \textit{ultra-local Carrollian flat bands}.
It is easy to see that this condition is satisfied whenever $\Lambda_{\mathbf k+\mathbf q,\mathbf k}$ depends only on the momentum transfer $\mathbf q$ and not on $\mathbf k$. In fact the converse is also true as we show now; when the condition is satisfied, $\Lambda_{\mathbf k+\mathbf q,\mathbf k}$ depends only on $\mathbf{q}$ in a suitable gauge.
The vanishing of the bracket of \eqref{eq:densityalgebra} for infinitesimal $\mathbf{q}^\prime$ implies, up to the lowest order in $\mathbf{q}^\prime$,
\begin{equation}
  \mathcal D_{k_i}\Lambda_{\mathbf k+\mathbf q,\mathbf k}\equiv
  \partial_{k_i}\Lambda_{\mathbf k+\mathbf q,\mathbf k}
  -\iu\big[\mathcal A_i(\mathbf k+\mathbf q)-\mathcal A_i(\mathbf k)\big]\Lambda_{\mathbf k+\mathbf q,\mathbf k}=0,
  \label{eq:dlog}
\end{equation}
where $\mathcal A_i\equiv \iu\langle u_{\mathbf k}|\partial_{k_i}u_{\mathbf k}\rangle$ is the Berry
connection of the flat band; $\mathcal D_{k_i}$ is the gauge-covariant momentum derivative of the form factor.
Since an
ultra-local Carrollian band is in particular a Carrollian flat band, $\Omega_{ij}(\mathbf k)=0$
holds. Then the Berry connection can be written as a gradient of some function $\varphi(\mathbf k)$ as
$\boldsymbol{\mathcal A}(\mathbf k)=\nabla_{\mathbf k}\varphi(\mathbf k)$ up to a constant, and
Eq.~\eqref{eq:dlog} integrates to
\begin{equation}
  \Lambda_{\mathbf k+\mathbf q,\mathbf k}=\eu^{\iu[\varphi(\mathbf k+\mathbf q)-\varphi(\mathbf k)]}\lambda(\mathbf q) \label{eq:lambda}
\end{equation}
for some function $\lambda(\mathbf q)$.
The phase can be removed by the gauge transformation $|u_{\mathbf k}\rangle\to \eu^{\iu\varphi(\mathbf k)}|u_{\mathbf k}\rangle$. In this gauge, $\Lambda_{\mathbf k+\mathbf q,\mathbf k} = \lambda(\mathbf{q})$, and thus the form factor depends on $\mathbf q$ alone. Related constraints on form factors follow from the classification of closed projected-density algebras~\cite{WangSimon2025}.

We note that an ultra-local Carrollian band is Carrollian but the converse is not true: a $\mathbf k$-independent form factor constrains all orders in $\mathbf q$, whereas $\Omega_{ij}(\mathbf{k}) = 0$ is a condition at the lowest order in $\mathbf{q}$ only.
In Supplemental Material, we connect this closure condition to microscopic hopping constraints and distinguish it from charge conservation alone.

We now argue that the ultra-local condition can be identified as the vanishing of the minimal quantum metric. The quantum metric~\cite{Provost1980,Resta2011,Torma2022} of the band is
\begin{equation}
  g_{ij}(\mathbf k)=\mathrm{Re}\,\langle\partial_{k_i}u_{\mathbf k}|
  \big(1-|u_{\mathbf k}\rangle\langle u_{\mathbf k}|\big)|\partial_{k_j}u_{\mathbf k}\rangle .
  \label{eq:metric}
\end{equation}
Since $|u_{\mathbf k}\rangle=\eu^{-\iu\mathbf k\cdot\mathbf x}|\psi_{\mathbf k}\rangle$, the metric depends on the positions $\boldsymbol\delta_a$ assigned to the orbitals, where $a$ labels each orbital. Shifting $\boldsymbol\delta_a\to\boldsymbol\delta_a+\boldsymbol\Delta_a$ multiplies the $a$-orbital component $u_a(\mathbf k)$ of $|u_{\mathbf k}\rangle$ by $\eu^{-\iu\mathbf k\cdot\boldsymbol\Delta_a}$ and generally changes $g_{ij}(\mathbf{k})$, known as the orbital embedding dependence~\cite{SimonRudner2020,Huhtinen2022}.
The minimal quantum metric $g^{\min}_{ij}(\mathbf k)$ is the metric in the embedding that minimizes $\int \mathrm{d}^2k\,\mathrm{tr}\,g(\mathbf k)$, where the embedding consists of a choice of orbital basis at each position, i.e. $\mathbf k$-independent unitary rotations among the orbitals sharing that position, followed by an independent shift of position $\boldsymbol\delta_a+\boldsymbol\Delta_a$ for each orbital of that basis. When each position hosts a single orbital, no orbital basis choice is necessary and this reduces to the definition of Ref.~\cite{Huhtinen2022}.
Since $g_{ij}(\mathbf{k})$ is positive semi-definite, the integral of its trace vanishes if and only if there is a choice of orbital embedding where $g_{ij}(\mathbf k)=0$ throughout the Brillouin zone. (We note that, throughout this paper, the Berry curvature $\Omega_{ij}(\mathbf{k})$ is always calculated with physical positions of orbitals.) We now show that the minimal quantum metric vanishes if and only if the band is ultra-local Carrollian. 

Suppose first that $g_{ij}(\mathbf k)=0$ for some embedding. 
Taking the trace of Eq.~\eqref{eq:metric}, $\mathrm{tr}\,g=\sum_i\big\|\big(1-|u_{\mathbf k}\rangle\langle u_{\mathbf k}|\big)|\partial_{k_i}u_{\mathbf k}\rangle\big\|^2=0$. Thus $|\partial_{k_i}u_{\mathbf k}\rangle = -\iu\mathcal{A}_i (\mathbf{k})|u_{\mathbf k}\rangle$, which integrates to $|u_{\mathbf k}\rangle=\eu^{-\iu\varphi(\mathbf k)}|c\rangle$ with $\nabla \varphi(\mathbf{k}) = \boldsymbol{\mathcal{A}}(\mathbf{k})$ and $|c\rangle$ independent of $\mathbf k$. Undoing the shift of positions, the Bloch function in
the original embedding is $u_a(\mathbf k)=\eu^{-\iu\varphi(\mathbf k)}c_a\,\eu^{\iu\mathbf k\cdot\boldsymbol\Delta_a}$.
Then, the form factor is
\begin{align}
    \Lambda_{\mathbf{k} + \mathbf{q},\mathbf{k}}
    =
    \eu^{\iu[\varphi(\mathbf{k}+\mathbf{q}) -\varphi(\mathbf{k})]}\sum_a \eu^{-\iu\mathbf{q}\cdot\boldsymbol\Delta_a} |c_a|^2.
\end{align}
This is exactly of the form \eqref{eq:lambda} and thus the phase factor $\eu^{\iu[\varphi(\mathbf{k}+\mathbf{q}) -\varphi(\mathbf{k})]}$ can be gauged away as before. Thus, the form factor is $\mathbf{k}$-independent in a suitable gauge and the band is ultra-local Carrollian. 

Conversely, suppose the projected densities commute. After a suitable $\mathbf k$-independent
rotation among orbitals at the same position, the band is spanned by the translates
$|v_{\mathbf{r}}\rangle=T_{\mathbf{r}}|v_{\mathbf 0}\rangle$ of a single compact localized
state~\cite{Sutherland1986,Flach2014,Leykam2018,RhimYang2019,Sathe2021} with mutually disjoint
supports (End Matter). Disjointness implies that no orbital label $a$ appears in
$|v_{\mathbf 0}\rangle$ in two different unit cells, $|v_{\mathbf 0}\rangle=\sum_ac_a|\mathbf{r}_a,a\rangle$,
so that the Bloch states $|\psi_{\mathbf k}\rangle=N^{-1/2}\sum_{\mathbf{r}}\eu^{\iu\mathbf k\cdot\mathbf{r}}|v_{\mathbf{r}}\rangle$
have components
\begin{equation}
  u_a(\mathbf k)=c_a\,\eu^{-\iu\mathbf k\cdot(\mathbf{r}_a+\boldsymbol\delta_a)} ,
  \label{eq:monomialmain}
\end{equation}
a constant times a single Bloch phase (End Matter); we call such bands \emph{monomial}.
Shifting orbital $a$ by $\boldsymbol\Delta_a=\mathbf d-\mathbf{r}_a-\boldsymbol\delta_a$, which
places all orbitals of $|v_{\mathbf 0}\rangle$ at one point $\mathbf d$, turns
Eq.~\eqref{eq:monomialmain} into $u_a(\mathbf k)=c_a\eu^{-\iu\mathbf k\cdot\mathbf d}$, for which
$g_{ij}(\mathbf k)=0$. Hence $g^{\min}=0$.

\section{Models}
Many known flat bands are (ultra-local) Carrollian. In one dimension there is no Berry
curvature, so every isolated flat band is Carrollian; what distinguishes them is
ultra-locality. The cross-stitch ladder \cite{Flach2014,Maimaiti2017}, whose
compact localized states occupy single rungs on a connected lattice, is ultra-local, whereas the sawtooth~\cite{Maimaiti2017,Leykam2018} and stub (one-dimensional
Lieb) chains~\cite{Real2017} are Carrollian but not ultra-local: their compact localized states overlap and $g^{\min}\neq0$. In two dimensions, combined time-reversal and inversion symmetry acting on the physical lattice, orbital positions included, forces the Berry curvature to vanish; Tasaki's decorated lattices \cite{Tasaki1992,Tasaki1998} are Carrollian but not ultra-local in this way. The Lieb lattice~\cite{Lieb1989,Julku2016} gapped by alternating hoppings breaks inversion and is the standard non-Carrollian example. (Singular flat bands \cite{RhimYang2019}, which are gapless, are outside our scope.)

Below we use the cross-stitch lattice and an extension of it. Each site $\mathbf{r}$ of a chain (one dimension) or of a triangular lattice (two dimensions) carries two orbitals, and the Hamiltonian is
\begin{align}
  H=&
  t_0\sum_{\mathbf{r}}\big(|\mathbf{r},1\rangle\langle\mathbf{r},2|+{\rm h.c.}\big)
  \notag \\
   &+t\sum_{\langle\mathbf{r}\mathbf{r}'\rangle}\sum_{a,b=1,2}\big(|\mathbf{r},a\rangle\langle\mathbf{r}',b|+{\rm h.c.}\big)
   \label{eq:crossstitch} \\
   =&
  2t_0\sum_{\mathbf{r}}|s_{\mathbf{r}}\rangle\langle s_{\mathbf{r}}|
   +2t\sum_{\langle\mathbf{r}\mathbf{r}'\rangle}\big(|s_{\mathbf{r}}\rangle\langle s_{\mathbf{r}'}|+{\rm h.c.}\big)
   -t_0, \notag
\end{align}
with $|s_{\mathbf{r}}\rangle=(|\mathbf{r},1\rangle+|\mathbf{r},2\rangle)/\sqrt2$ being the symmetric combination of the two orbitals of a site. 
(In the numerical simulation below, we set $t_0=4$, $t=1/2$.)
Since $H+t_0$ involves only the symmetric orbitals, the antisymmetric combinations $|a_{\mathbf{r}}\rangle=(|\mathbf{r},1\rangle-|\mathbf{r},2\rangle)/\sqrt2$ are exact eigenstates at $E_0=-t_0$: a flat band of single-site compact localized states with disjoint supports, i.e. an ultra-local Carrollian flat band, while the symmetric orbitals form a dispersive band separated by a gap for $t_0>2|t|$ (chain) or $t_0>3t$ for $t > 0$ (triangular lattice).
We extend this model by replacing $|s_{\mathbf{r}}\rangle$ in Eq.~\eqref{eq:crossstitch} with a dressed orbital carrying a tail on the neighboring sites,
\begin{equation}
  |s_{\mathbf{r}}\rangle\;\to\; \frac{1}{\sqrt{2}}\bigg(|\mathbf{r},1\rangle+|\mathbf{r},2\rangle+\tau\sum_{\mathbf d}|\mathbf{r}+\mathbf d,1\rangle\bigg)\,,
  \label{eq:dressed}
\end{equation}
where $\mathbf d$ runs over the nearest-neighbor vectors. The flat band is the set of states
orthogonal to all dressed orbitals, which forces the compact localized states to acquire a tail too,
$|\mathbf{r},1\rangle-|\mathbf{r},2\rangle-\tau\sum_{\mathbf d}|\mathbf{r}+\mathbf d,2\rangle$, and the band stays exactly flat at $-t_0$. The compact localized states are real and inversion symmetric, so the Berry curvature vanishes and the band is Carrollian. Neighboring compact localized states now overlap, however, and the band is no longer ultra-local, with $\langle\mathrm{tr}\,g^{\min}\rangle=\tfrac32\tau^2+O(\tau^3)$ on the triangular lattice.
This is the cell construction of Tasaki~\cite{Tasaki1992,Tasaki1998,MielkeTasaki1993}; in the orbital basis the extension modifies the nearest-neighbor hoppings and adds second- and third-neighbor ones with amplitudes fixed by
$\tau$. Giving the tail to only three of the six neighbors of the triangular lattice instead breaks inversion and yields a non-Carrollian flat band with nonzero Berry curvature (End Matter).

\section{Hierarchy of Immobility}
\label{sec:immobility}

The word ``Carrollian'' promises that nothing moves, and the two conditions above turn this into
a hierarchy. Consider $N_p$ particles confined to the flat band. An overbar denotes the projection
of a single-particle operator, $\bar A=PAP$, and a tilde denotes the
corresponding projected operator in the $N_p$-particle flat-band Hilbert space,
$\widetilde A \equiv \sum_{\alpha=1}^{N_p}(\bar A)_\alpha$
where $\alpha$ labels the particle on which the operator acts. For a many-body
operator, we use the same tilde to denote projection onto the many-body flat-band Hilbert
space; explicitly, $\widetilde A=\mathcal P A\mathcal P$, where $\mathcal P$ is the
$N_p$-particle projector associated with $P$.

The center of mass is described by the projected dipole
\begin{align}
    \tilde{D}_i = \sum_{\alpha = 1}^{N_p} (P x_i P)_\alpha.
\end{align}
We subject the particles to a uniform force $\mathbf F$ and to a translationally
invariant two-body interaction $V=\frac{1}{2}\sum_{\alpha\neq\beta}V(\mathbf x_\alpha-\mathbf x_\beta)$.
The projected many-body Hamiltonian is
\begin{align}
    \widetilde H
    &=
    E_0N_p
    -
    \mathbf F\cdot\widetilde{\mathbf D}
    +
    \widetilde V,
    \label{eq:Hproj}
    \\
    \widetilde V
    &\equiv
    \mathcal P V\mathcal P
    =
    \frac{1}{2}
    \sum_{\mathbf q}
    V(\mathbf q)
    \sum_{\alpha\neq\beta}
    (\bar n_{\mathbf q})_\alpha
    (\bar n_{-\mathbf q})_\beta.
    \label{eq:Vprojected}
\end{align}
with the single-particle projected density $\bar n_{\mathbf q}$ defined in
Eq.~\eqref{eq:nbar}. The time evolution of the projected dipole is
\begin{align}
    \frac{\mathrm{d}}{\mathrm{d}t}\langle\widetilde D_i\rangle
    &=
    \iu\langle[\widetilde H,\widetilde D_i]\rangle
    =
    -F_j\langle\widetilde\Omega_{ij}\rangle
    +
    \iu\langle[\widetilde V,\widetilde D_i]\rangle ,
    \label{eq:velocity}
\end{align}
where $\widetilde\Omega_{ij}$ is the many-body counterpart of the single-particle Berry-curvature operator
$\bar\Omega_{ij}
    \equiv
    \sum_{\mathbf k}
    \Omega_{ij}(\mathbf k)
    |\psi_{\mathbf k}\rangle
    \langle\psi_{\mathbf k}|
$.
The first term in Eq.~\eqref{eq:velocity} is the anomalous velocity~\cite{Karplus1954,Chang1996,Sundaram1999,Xiao2010}, and it vanishes identically for a Carrollian flat band.

\begin{figure*}[t]
\includegraphics[width=0.66\columnwidth]{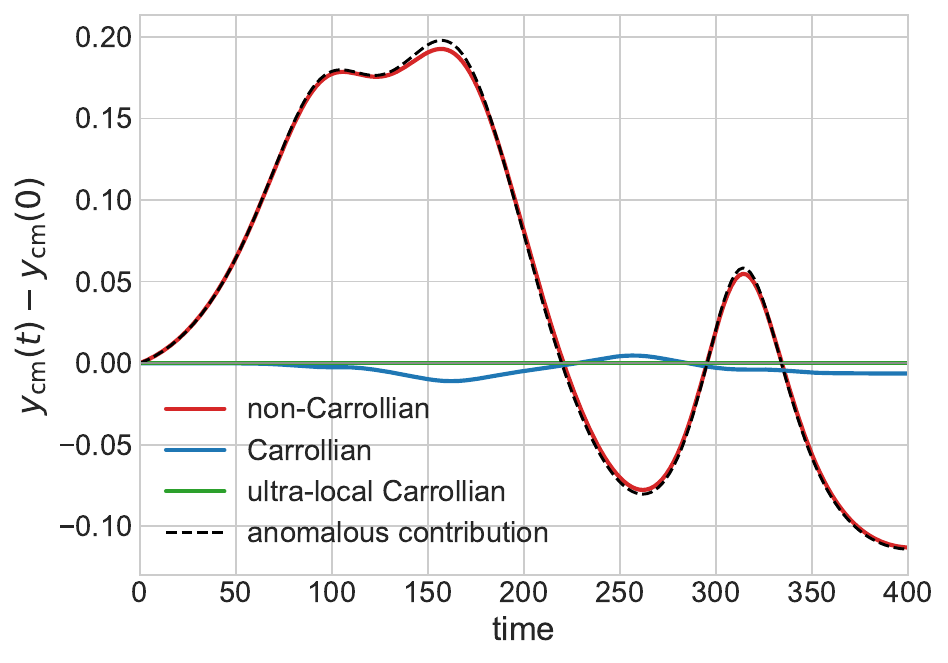}
\includegraphics[width=0.66\columnwidth]{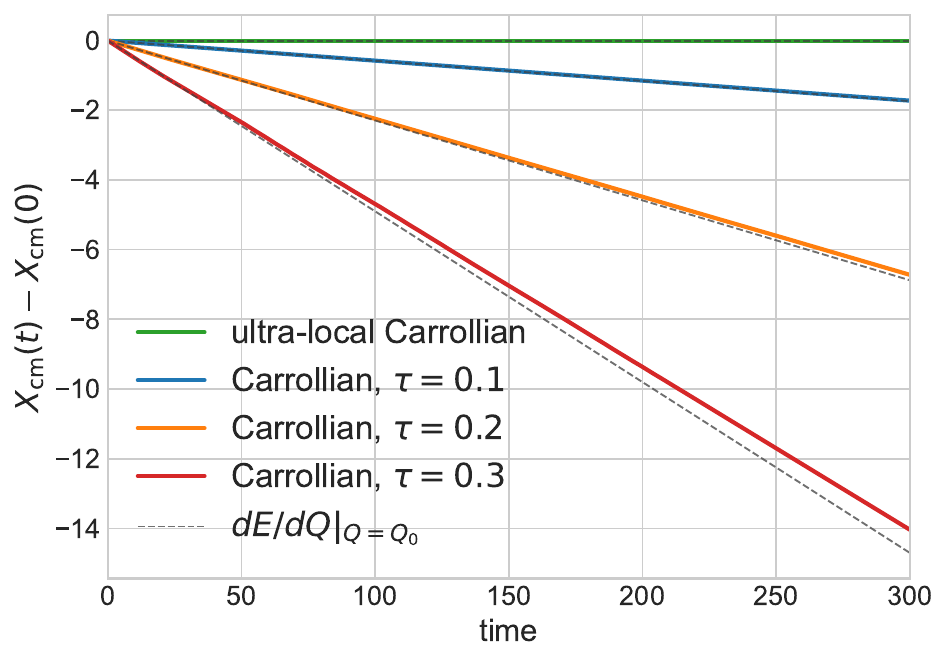}
\includegraphics[width=0.66\columnwidth]{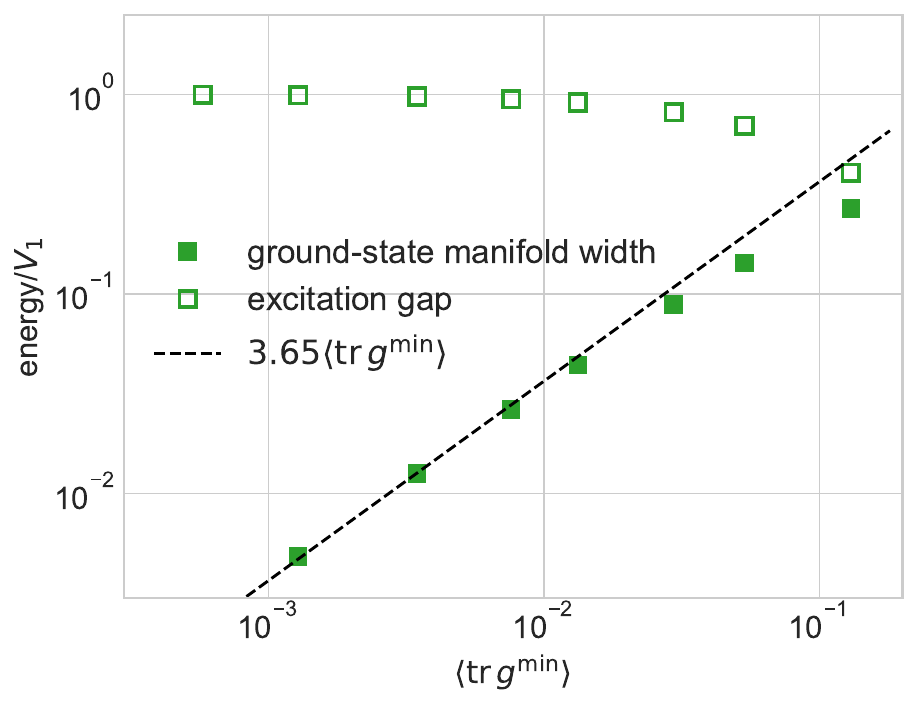}
\caption{Hierarchy of immobility. (a) Transverse center-of-mass displacement of a non-interacting
wave packet (width $\sigma = 4$ and momentum $\mathbf{k}_0 = (0,-\pi/(2\sqrt{3}))$) on the flat band of the triangular-lattice cross-stitch model ($30 \times 30$ sites) under a uniform force $F=-0.03$ along the $x$-direction, for the non-Carrollian, Carrollian ($\tau=0.5$) and ultra-local Carrollian bands. Dashed: semiclassical anomalous displacement $F\int_0^t \mathrm{d}t'\,\Omega_{xy}\left(k_{0x}+Ft',k_{0y}\right)$. (b) Center of mass of two fermions on the open chain of length 100 sites with nearest-neighbor repulsion $V_1 = 1$.
The initial state is a Gaussian wave packet of two fermions occupying neighboring compact localized states, with center-of-mass width $\sigma=6$ and total momentum $Q_0=\pi/2$. The pair is immobile on the ultra-local band ($\tau=0$) and moves at close to the group velocity $\mathrm{d}E/\mathrm{d}Q|_{Q = Q_0}$ of the interaction-generated bound-pair dispersion (dashed) on the Carrollian bands.
(c) Exact diagonalization of $N_p = 4$ spinless fermions with nearest-neighbor repulsion $V_1$ on a $3 \times 6$ cross-stitch lattice on a triangular torus. The width of the energy of the ground state manifold (filled) and the excitation gap (open) versus the Brillouin-zone-averaged minimal metric $\langle\mathrm{tr}\,g^{\min}\rangle$. The dashed line $\propto \langle\mathrm{tr}\,g^{\min}\rangle$ is a linear fit at small $\langle\mathrm{tr}\,g^{\min}\rangle$ to show the scaling.
}
\label{fig:immobility}
\end{figure*}

The second term in Eq.~\eqref{eq:velocity} is the motion induced by interactions. The relevant single-particle
commutator is
\begin{equation}
    [\bar n_{\mathbf q}, Px_iP]
    =
    -\iu\sum_{\mathbf k}
    \big(
    \mathcal D_{k_i}
    \Lambda_{\mathbf k+\mathbf q,\mathbf k}
    \big)
    |\psi_{\mathbf k+\mathbf q}\rangle
    \langle\psi_{\mathbf k}|,
    \label{eq:DLambda}
\end{equation}
where $\mathcal D_{k_i} \Lambda_{\mathbf k+\mathbf q,\mathbf k}$ is the covariant derivative defined in \eqref{eq:dlog}. The gauge-covariant momentum dependence of the form factor thus controls the motion generated by interactions.
For a Carrollian band, where the Berry curvature vanishes, at small $\mathbf q$,
\begin{align}
  \mathcal D_{k_i}\Lambda_{\mathbf k+\mathbf q,\mathbf k}=-\tfrac12\,q_aq_b\,\partial_{k_i}g_{ab}(\mathbf k)+O(q^3).
\end{align}
so the leading interaction-induced motion is controlled by the momentum dependence of the quantum metric. Related manifestations of quantum geometry in interaction-enabled flat-band transport include the geometric contribution to the superfluid weight and the effective mass of two-body bound states~\cite{Peotta2015,Torma2018}.
For an ultra-local Carrollian band, Eq.~\eqref{eq:dlog} implies $[\widetilde V,\widetilde D_i]=0$ for any translationally invariant two-body interaction.

Equation~\eqref{eq:velocity} thus separates three classes. In a generic flat band the group velocity vanishes, but a wave packet moves under a force, transversely to it, through the anomalous velocity.
In a Carrollian flat band, a uniform force alone leaves the center of mass of any state unchanged within the band, which is the physical content of $[\bar B_i,\bar B_j]=0$, since the force term $-\mathbf F\cdot\widetilde{\mathbf D}=-(\mathbf F/E_0)\cdot\widetilde{\mathbf B}$ is itself a boost generator. Interactions and non-uniform one-body potentials $W(\mathbf x)$ couple instead to the supertranslation generators, $PWP=\sum_{\mathbf q}\tilde W(\mathbf q)\bar n_{\mathbf q}$, and can still displace particles through $\mathcal D_{k_i}\Lambda$, by an amount controlled by
the quantum metric. Interactions thus break the conservation of the Carroll boosts in a Carrollian band that is not ultra-local. In an ultra-local Carrollian band both terms vanish, and so does the motion caused by any one-body potential.

We now numerically verify this hierarchy.
Fig.~\ref{fig:immobility}(a) shows a wave packet motion prepared in the flat band of the triangular cross-stitch lattice and its extensions to non-ultra-local Carrollian and non-Carrollian flat bands. We apply a uniform force along $x$ and evolve in time with the full two-band Hamiltonian. The non-Carrollian band moves transversely, in agreement with the anomalous velocity, while the Carrollian and ultra-local Carrollian bands do not. The small residual displacement of the Carrollian band is of higher order in force, which comes from the inter-band mixing. Fig.~\ref{fig:immobility}(b) shows two interacting fermions on the cross-stitch chain, prepared as a pair wave packet with center-of-mass momentum $Q_0 = \pi/2$. As previously studied in Refs.~\cite{Vidal2000,Tovmasyan2018,Martinez2023,Chen2025}, interactions can make particles move even in a flat band due to the quantum metric~\cite{Tovmasyan2018,Torma2018}. We observe that, on the ultra-local Carrollian band, there is no motion even in the presence of interactions, whereas on the non-ultra-local Carrollian bands the center of mass moves at close to $\mathrm{d}E/\mathrm{d}Q|_{Q = Q_0}$, where $E(Q)$ is the energy of the bound pair at total momentum $Q$.

The last step of the hierarchy has a direct many-body consequence. On an ultra-local band, every eigenstate
can be chosen to be a configuration of occupied compact localized states and the interacting flat band is exactly a classical lattice gas at every filling and temperature, for any density-density interaction, with a coupling $U$ obtained by averaging the bare interaction over the shapes of two compact localized states.
The fact that the flat bands built from non-overlapping localized states map to classical lattice gas has been recognized before~\cite{SchulenburgRichter2002,ZT2004,DRM2015}, and suitable interactions in flat-band lattices lead to an extensive set of local integrals of motion~\cite{Danieli2020}.
For the triangular cross-stitch lattice with nearest-neighbor repulsion $V_1$, we have $U = V_1$ and the lattice gas is Baxter's hard-hexagon model~\cite{Baxter1980,Baxter1982} for $T\ll V_1$, with its exactly known close packing at $\nu=1/3$ and crystallization at $\nu_c=(5-\sqrt5)/10$.

Away from the ultra-local point the corrections are controlled by the minimal quantum metric. In the basis of maximally localized Wannier functions of the flat band~\cite{Marzari1997,Marzari2012}, the off-diagonal elements of the projected densities are at most of order the tail amplitude $\eta$ of a Wannier function on a neighboring site, with $\eta^2\propto\langle\mathrm{tr}\,g^{\min}\rangle$. Both the correlated hops they generate and the longer-range classical coupling contribute corrections of order $V_1\langle\mathrm{tr}\,g^{\min}\rangle$ within the zero-energy classical ground-state manifold (End Matter), while the gap from the ground-state manifold to genuine excitations remains $\sim V_1$. Fig.~\ref{fig:immobility}(c) confirms that the splitting of the ground-state manifold is proportional to the minimal quantum metric.

\section{Conclusions}

Carrollian symmetry is realized on a flat band if and only if its Berry curvature vanishes, and
its ultra-local extension if and only if its minimal quantum metric vanishes, which singles out
the monomial bands built from compact localized states with disjoint supports. These conditions
define a hierarchy of immobility, whose last step makes the interacting problem exactly a classical lattice gas, with corrections of order $V_1\langle\mathrm{tr}\,g^{\min}\rangle$ away from it.
Flat-band lattices have already been realized in cold-atom and photonic platforms~\cite{Taie2015,Mukherjee2015,Vicencio2015,Leykam2018}, so that two-orbital constructions of the kind studied here should be experimentally accessible.
Carrollian symmetry has striking consequences in high-energy physics, from ultra-local dynamics
to infinite-dimensional symmetry algebras~\cite{Henneaux1979,Duval2014bms,deBoer2022}; flat bands with the geometric properties identified here offer a place to realize them in the laboratory.

\begin{acknowledgments}
\textit{Acknowledgments}---
We used Claude Opus 5.5,  Claude Fable 5.1, and GPT-6.2 Astra to discuss and test ideas, check derivations, edit text, assist with numerical codes, and format the bibliography. All results were verified and all the text and bibliography were checked and revised by the authors, who take full responsibility for the content. We are supported by JSPS KAKENHI Grant No. JP24K00548, JP24K00634, JST PRESTO Grant No. JPMJPR2353, JST CREST Grant No. JPMJCR26XA, Frontier Research in Duo (FRiD) grant from Tohoku University, and JST BOOST Program Japan Grant No. JPMJBY24E0.\\
\end{acknowledgments}

\bibliographystyle{apsrev4-2}
\bibliography{bibliography}

\section{End Matter}
\subsection{Carrollian Symmetry and Supertranslations}

Carrollian symmetry was first identified as the symmetry arising from the vanishing-speed-of-light limit, $c \to 0$, of Poincaré symmetry, in contrast to the Galilean symmetry obtained in the non-relativistic limit $c \to \infty$. In \(d\) spatial dimensions, the Carroll group consists of time translations, spatial translations, spatial rotations, and Carrollian boosts. The latter act on spacetime coordinates as
\begin{align}
t \to t-\boldsymbol{\beta}\cdot\mathbf{x}\,,
\qquad \text{and} \qquad 
\mathbf{x}\to\mathbf{x}\,,
\end{align}
where \(\boldsymbol{\beta}\) is a constant Carrollian boost parameter. This is in contrast to Galilean (non-relativistic) boosts, which act as $t \to t$ and $\mathbf{x} \to \mathbf{x} - \mathbf{v} t$, where $\mathbf{v}$ is a constant Galilean boost parameter. 

Denoting the generators of time translations, spatial translations, rotations, and Carrollian boosts by $H$, $P_i$, $J_{ij}$, and $B_i$, respectively, with $i = 1, 2, \dots, d$, the nontrivial commutation relations of the global Carroll algebra are
\begin{equation}
\begin{aligned}
\relax [J_{ij},J_{kl}]&=\mathrm{i} (\delta_{ik}J_{jl}-\delta_{il}J_{jk}
-\delta_{jk}J_{il}+\delta_{jl}J_{ik}) \,,\\
[J_{ij},P_k]&= \mathrm{i} (\delta_{ik}P_j-\delta_{jk}P_i) \,,\\
[J_{ij},B_k]&= \mathrm{i} (\delta_{ik}B_j-\delta_{jk}B_i) \,,\\
[B_i,P_j]&=\mathrm{i} \delta_{ij}H \,.
\end{aligned}
\end{equation}
In particular, the Carrollian boosts commute among themselves and with the time-translation Hamiltonian,
\begin{align}
[B_i,B_j]=0,\qquad \text{and}\qquad [B_i,H]=0.
\end{align}

An important feature of Carrollian systems is the possible infinite-dimensional extension of the global Carrollian symmetry to arbitrary position-dependent time translations, known as \emph{Carrollian supertranslations},
\begin{align}
t\to t-f(\mathbf{x}) \,, \qquad \text{and} \qquad
\mathbf{x}\to\mathbf{x} \,,
\end{align}
where $f(\mathbf{x})$ is an arbitrary function of the spatial coordinates. Denoting their generators by $Q_f$, the supertranslations form an abelian algebra,
\begin{align}
[Q_f,Q_g]=0 \,.
\end{align}
Global time translations and Carrollian boosts correspond to $f(\mathbf{x})= \text{constant}$ and $f(\mathbf{x})=\boldsymbol{\beta}\cdot\mathbf{x}$, respectively. The infinite-dimensional extension therefore contains the global Carrollian symmetry as a subalgebra. While such an extension is characteristic of ultra-local Carrollian theories, it is not required by the global Carroll algebra alone. As demonstrated in the main text, this distinction naturally separates Carrollian flat bands from the more restrictive ultra-local Carrollian flat bands.

\subsection{Quantum geometry of ultra-local Carrollian flat bands}
Here we show that when the band is ultra-local Carrollian, i.e.\ the projected densities commute, the minimal quantum metric is zero.

We define the total density at position $\mathbf{r}$, summed over the orbitals located there, as $N_{\mathbf{r}}\equiv\sum_{a:\,\mathbf{r}+\boldsymbol\delta_a=\mathbf{r}}|\mathbf{r},a\rangle\langle\mathbf{r},a|$ for the orbital $a$ in the unit cell at lattice vector $\mathbf{r}$. Its projection onto the flat band is $\bar N_{\mathbf{r}}\equiv PN_{\mathbf{r}}P$.
Since $I = \sum_\mathbf{r}N_\mathbf{r}$, we can write $\eu^{\iu\mathbf q\cdot\mathbf x}=\sum_{\mathbf{r}}\eu^{\iu\mathbf q\cdot\mathbf{r}}N_{\mathbf{r}}$, and the projected density of Eq.~\eqref{eq:nbar} is $\bar n_{\mathbf q}=\sum_{\mathbf{r}}\eu^{\iu\mathbf q\cdot\mathbf{r}}\bar N_{\mathbf{r}}$, and the ultra-local Carrollian condition, Eq.~\eqref{eq:densityalgebra} for all $\mathbf q,\mathbf q'$, is equivalent to $[\bar N_{\mathbf{r}},\bar N_{\mathbf{r}'}]=0$ for all pairs of positions. The commuting operators $\bar N_{\mathbf{r}}$ have a common orthonormal eigenbasis $\{|w_j\rangle\}$ of the band, $\bar N_{\mathbf{r}}|w_j\rangle=\lambda_{\mathbf{r},j}|w_j\rangle$, with $\lambda_{\mathbf{r},j}=\langle w_j|N_{\mathbf{r}}|w_j\rangle$ and $\sum_{\mathbf{r}}\lambda_{\mathbf{r},j}=1$.
A nonzero $\lambda_{\mathbf{r},j}$ is a nonzero eigenvalue of $\bar N_{\mathbf{r}}=PN_{\mathbf{r}}(PN_{\mathbf{r}})^\dagger$, hence also a nonzero eigenvalue of $(PN_{\mathbf{r}})^\dagger PN_{\mathbf{r}} = N_{\mathbf{r}} P N_{\mathbf{r}}$.
By translation invariance, $N_{\mathbf{r}}PN_{\mathbf{r}}$ and $N_{\mathbf{r} + \mathbf{R}}PN_{\mathbf{r} + \mathbf{R}}$ are unitarily related, and hence this matrix takes only finitely many distinct forms, so their nonzero eigenvalues have a positive lower bound $\mu_{\min}$. Since $\sum_{\mathbf{r}}\lambda_{\mathbf{r},j}=1$, each $|w_j\rangle$ is nonzero on at most $1/\mu_{\min}$ positions.
For a position $\mathbf{r}$ with $\lambda_{\mathbf{r},j}\neq0$, the eigenvalue equation gives $|w_j\rangle=\bar N_{\mathbf{r}}|w_j\rangle/\lambda_{\mathbf{r},j}$, which lies in the span of the projected orbitals at $\mathbf{r}$. These are exponentially localized for an isolated band~\cite{Kohn1959,desCloizeaux1964,Brouder2007,Panati2007}, so all positions where $|w_j\rangle$ is nonzero lie within a bounded distance of $\mathbf{r}$. Each joint eigenspace, which is the span of the $|w_j\rangle$ with a given set of eigenvalues $\{\lambda_{\mathbf{r},j}\}_{\mathbf{r}}$, is thus finite dimensional with bounded support, and a nonzero lattice translation maps it to a different one. Therefore, the basis $\{|w_j\rangle\}$ can be chosen covariant under translations, namely translating any $|w_j\rangle$ by a lattice vector gives another basis state, with its support translated accordingly. For $i\neq j$,
$\langle w_i|N_{\mathbf{r}}|w_j\rangle=\langle w_i|\bar N_{\mathbf{r}}|w_j\rangle=0$.
Thus the parts $N_{\mathbf{r}}|w_j\rangle$ of different $|w_j\rangle$ at any one position $\mathbf{r}$ are mutually orthogonal vectors in the space of orbitals at $\mathbf{r}$. We take these vectors, normalized and completed to an orthonormal basis of that space if necessary, as the new orbitals at $\mathbf{r}$. By translation covariance this $\mathbf k$-independent rotation is the same in every unit cell, and it is trivial when each position hosts a single orbital. Below $|\mathbf{r},a\rangle$ denotes the rotated orbitals.
Each overlaps at most one $|w_j\rangle$, so $\bar n_{\mathbf{r}a}\equiv P|\mathbf{r},a\rangle\langle\mathbf{r},a|P\propto|w_j\rangle\langle w_j|$ has rank at most one and all $\bar n_{\mathbf{r}a}$ commute. Hence the projected orbitals $|v_{\mathbf{r}a}\rangle\equiv P|\mathbf{r},a\rangle$ are pairwise orthogonal or parallel.

The states $|w_j\rangle$ are thus orthonormal eigenstates of $H$ with bounded and, in the rotated
basis, mutually disjoint orbital supports: compact localized states spanning the band.
Since the basis is covariant under translations, the $N$ translates $T_{\mathbf{r}}|w_j\rangle$ of any one of
them are again basis states; they are distinct, because a bounded support is never invariant under
a nonzero translation, and since the rank of $P$ is $N$, they exhaust the band. 
Choosing any one of the $|w_j\rangle$ as $|v_{\mathbf 0}\rangle$, every basis state is one of its
translates, and we label the compact localized states by lattice vectors,
$|v_{\mathbf{r}}\rangle=T_{\mathbf{r}}|v_{\mathbf 0}\rangle$, with support $S_{\mathbf{r}}$ obtained
from $S_{\mathbf 0}$ by translating every position by $\mathbf{r}$. 

The cell-periodic part of the Bloch state $|u_\mathbf{k}\rangle$ for ultra-local Carrollian bands takes a particularly simple form.
Disjointness of the supports implies that each orbital label $a$ appears in
$|v_{\mathbf 0}\rangle$ at most once: if it appeared at two positions
$\mathbf{r}_a+\boldsymbol\delta_a$ and $\mathbf{r}'_a+\boldsymbol\delta_a$, the translate of
$|v_{\mathbf 0}\rangle$ by $\mathbf{r}'_a-\mathbf{r}_a$ would share a position with
$|v_{\mathbf 0}\rangle$. Hence $|v_{\mathbf 0}\rangle=\sum_ac_a|\mathbf{r}_a,a\rangle$ with
$\sum_a|c_a|^2=1$ and $c_a=0$ outside the support, and
$|v_{\mathbf{r}}\rangle=\sum_ac_a|\mathbf{r}+\mathbf{r}_a,a\rangle$. For the Bloch states
$|\psi_{\mathbf k}\rangle=N^{-1/2}\sum_{\mathbf{r}}\eu^{\iu\mathbf k\cdot\mathbf{r}}|v_{\mathbf{r}}\rangle$,
\begin{align}
  &u_a(\mathbf k)
  =\sqrt N\,\langle\mathbf{r},a|\eu^{-\iu\mathbf k\cdot\mathbf x}|\psi_{\mathbf k}\rangle
  \\
  &=\eu^{-\iu\mathbf k\cdot(\mathbf{r}+\boldsymbol\delta_a)}\sum_{\mathbf{r}'}\eu^{\iu\mathbf k\cdot\mathbf{r}'}\langle\mathbf{r},a|v_{\mathbf{r}'}\rangle
  =c_a\,\eu^{-\iu\mathbf k\cdot(\mathbf{r}_a+\boldsymbol\delta_a)},\notag
\end{align}
since only $\mathbf{r}'=\mathbf{r}-\mathbf{r}_a$ contributes to the sum. This is
Eq.~\eqref{eq:monomialmain}, from which $g^{\min}=0$ follows as in the main text; the shift
$\boldsymbol\Delta_a$ there places all orbitals of the compact localized state at a single
point, making the generalized atomic insulator~\cite{Bradlyn2017} literally atomic.

\subsection{Minimal quantum metric of the extended cross-stitch lattice}

The flat-band state of the extended model, Eq.~\eqref{eq:dressed}, is real in the orbital
basis, $|w(\mathbf k)\rangle=(-1,\,1+\tau F(\mathbf k))^T$ with
$F(\mathbf k)=\sum_{\mathbf d}\cos\mathbf k\cdot\mathbf d$ summed over the nearest-neighbor
vectors. For a real normalized state $\langle u_{\mathbf k}|\partial_{k_i}u_{\mathbf k}\rangle=0$,
so that $g_{ij}=\langle\partial_{k_i}u_{\mathbf k}|\partial_{k_j}u_{\mathbf k}\rangle$, and for the normalized state $|u_{\mathbf k}\rangle=|w(\mathbf k)\rangle/\|w(\mathbf k)\|$ one finds
\begin{equation}
  g_{ij}(\mathbf k)=\frac{\tau^2\,\partial_{k_i}F\,\partial_{k_j}F}{\big[1+(1+\tau F)^2\big]^2}
  =\frac{\tau^2}{4}\,\partial_{k_i}F\,\partial_{k_j}F+O(\tau^3).
\end{equation}
With three independent neighbor vectors $\mathbf d_j$ of unit length, the Brillouin-zone
average $\langle\sin(\mathbf k\cdot\mathbf d_j)\sin(\mathbf k\cdot\mathbf d_l)\rangle=\delta_{jl}/2$
gives $\langle|\nabla_{\mathbf k}F|^2\rangle=6$ on the triangular lattice and $2$ on the chain,
hence $\langle\mathrm{tr}\,g\rangle=\tfrac32\tau^2+O(\tau^3)$ and $\tfrac12\tau^2+O(\tau^3)$,
respectively, in the physical embedding.

The minimization over embeddings does not lower this value at order $\tau^2$. A
$\mathbf k$-independent rotation of the two orbitals leaves $g_{ij}(\mathbf k)$ unchanged, and a
subsequent relative shift $\boldsymbol\Delta$ of the two orbitals adds
$-\iu\Delta_in_2|u_{\mathbf k}\rangle$ to $\partial_{k_i}|u_{\mathbf k}\rangle$, with
$n_2=|2\rangle\langle2|$ in the rotated basis. To order $\tau^2$ the resulting change of
$\mathrm{tr}\,g$ consists of a non-negative term proportional to $|\boldsymbol\Delta|^2$ and of
cross terms proportional to $\partial_{k_i}F$ or $\partial_{k_i}F^2$, whose Brillouin-zone
averages vanish. Hence $\langle\mathrm{tr}\,g^{\min}\rangle=\tfrac32\tau^2+O(\tau^3)$ on the
triangular lattice: the tail of the compact localized state sits on orbital $2$ of the
neighboring sites, which carries the same label as orbital $2$ of the central site, so that no
assignment of positions to orbital labels can bring these positions together.

\subsection{Non-Carrollian cross-stitch lattice}
With the primitive vectors $\mathbf a_1=(1,0)$ and $\mathbf a_2=(1,\sqrt3)/2$, the
non-Carrollian band used in Fig.~\ref{fig:immobility}(a) is obtained by restricting the tail
in Eq.~\eqref{eq:dressed} to $\mathbf d\in\{\mathbf a_1,\mathbf a_2,\mathbf a_2-\mathbf a_1\}$,
excluding the three opposite directions. This breaks inversion symmetry, and the Berry
curvature, odd under $\mathbf k\to-\mathbf k$ by time reversal, is nonzero. The compact
localized states then carry the tail in the opposite directions,
$|\mathbf{r},1\rangle-|\mathbf{r},2\rangle-\tau\sum_{\mathbf d}|\mathbf{r}-\mathbf d,2\rangle$.

\subsection{Corrections away from the ultra-local point}
For a band that is Carrollian but not ultra-local, the projected densities no longer commute
and the classical description of the classical lattice gas of the main text acquires corrections. The splitting of
the classical ground-state manifold is a property of the spectrum of $\widetilde V$ and does not
depend on the basis chosen for the flat band, but the decomposition of $\widetilde V$ into a
classical part and correlated hops does. We use the basis of maximally localized Wannier
functions of the flat band~\cite{Marzari1997,Marzari2012}, $P=\sum_{\mathbf{r}}|W_{\mathbf{r}}\rangle\langle W_{\mathbf{r}}|$. Near the ultra-local point each $|W_{\mathbf{r}}\rangle$ consists
of the compact localized state $|v_{\mathbf{r}}\rangle$ with amplitudes of order one and a tail of
amplitude $\eta$ on the neighboring compact localized states. The projected density at a site,
$\bar n_{\mathbf x}=\sum_{\mathbf{r},\mathbf{r}'}W_{\mathbf{r}}^*(\mathbf x)W_{\mathbf{r}'}(\mathbf x)|W_{\mathbf{r}}\rangle\langle W_{\mathbf{r}'}|$ with $W_{\mathbf{r}}(\mathbf x)=\langle\mathbf x|W_{\mathbf{r}}\rangle$, therefore has diagonal elements $|W_{\mathbf{r}}(\mathbf x)|^2$ that reproduce the classical couplings, with corrections of order $\eta^2$ from the tail weight, and off-diagonal elements $\mathbf{r}\neq\mathbf{r}'$ at most of order $\eta$, since the supports of two compact localized states are disjoint and a site can only be in the core of one Wannier function and in the tail of another.

Inserted into $\widetilde V$, the off-diagonal elements can generate correlated hops of amplitude $\sim V_1\eta$, in which a particle moves by one compact localized state while another particle is nearby.
A hop that creates a pair of neighboring occupied states leaves the ground-state manifold at a cost $V_1$ and contributes $\sim(V_1\eta)^2/V_1=V_1\eta^2$ in second-order perturbation theory; hops of amplitude $\sim V_1\eta^2$ that stay inside the manifold, in which the shared site is in the tail of both Wannier functions, and the longer-range classical couplings of order $V_1\eta^2$ from the tail weight contribute at first
order. 
All corrections to the classical spectrum are thus of order $V_1\eta^2$, while the gap to
genuine excitations remains of order $V_1$.

Finally, $\eta^2\propto\langle\mathrm{tr}\,g^{\min}\rangle$ near the ultra-local point. The
Brillouin-zone average of $\mathrm{tr}\,g$ is the gauge-invariant part of the spread of the
Wannier functions~\cite{Marzari1997}, which is quadratic in the tails. Its minimization over
orbital positions removes the contribution of the internal structure of the compact localized
state, which is present even when $\eta=0$, so that the minimal metric vanishes exactly when
the tails do and is quadratic in them, as verified for the present deformation by the metric calculation above. The
corrections to the classical lattice gas are therefore of order $V_1\langle\mathrm{tr}\,g^{\min}\rangle$,
the combination plotted in Fig.~\ref{fig:immobility}(c).

\clearpage

\begin{widetext}
\begin{center}
\Large{Supplemental Material for\\ ``Carrollian Flat Bands: Quantum Geometry and a Hierarchy of Immobility''}
\end{center}
\end{widetext}

\section{Extensions of the Carroll algebra}
When $\Omega_{ij}(\mathbf k)\neq0$, the projected boosts $\bar B_i=E_0Px_iP$ of Eq.~(1) still commute with $H$ and satisfy $T_{\mathbf a}^\dagger\bar B_iT_{\mathbf a}=\bar B_i+a_i\bar{H}$, but
\begin{equation}
[\bar B_i,\bar B_j]=\iu E_0^2\,\bar\Omega_{ij},
\end{equation}
with $\bar\Omega_{ij}=\sum_{\mathbf k}\Omega_{ij}(\mathbf k)|\psi_{\mathbf k}\rangle\langle\psi_{\mathbf k}|$ the Berry-curvature operator defined below Eq.~(16).
The operator $\bar\Omega_{ij}$ commutes with $H$ and with lattice translations, but not with the boosts
\begin{equation}
[\bar B_l,\bar\Omega_{ij}]=\iu E_0\sum_{\mathbf k}\partial_{k_l}\Omega_{ij}(\mathbf k)\,|\psi_{\mathbf k}\rangle\langle\psi_{\mathbf k}| ,
\end{equation}
because $\bar B_l$ acts on the coefficient of $|\psi_{\mathbf k}\rangle$ as $E_0(\iu \partial_{k_l}+\mathcal A_l)$ and $\bar\Omega_{ij}$ multiplies it by $\Omega_{ij}(\mathbf k)$.

Two cases should be distinguished. If $\Omega_{xy}$ is uniform, $\bar\Omega_{xy}=\Omega_{xy}P$ is a $c$-number on the band, and $[\bar B_x,\bar B_y]=\iu E_0^2\Omega_{xy}P$ is a central extension. In $d=2$ the Carroll algebra admits such a central extension of the boost commutator, which makes the boosts noncommutative and leads to Hall-type motion~\cite{Marsot2022,Marsot2023}. The band then carries a projective representation of the planar Carroll algebra, as the lowest Landau level does, and the anomalous velocity of the first step of the hierarchy is the lattice counterpart of this Hall motion.

A uniform and nonzero $\Omega_{xy}$, however, requires a nonzero Chern number, since $\int \mathrm{d}^2k\,\Omega_{xy}=2\pi C$, and an exactly flat band with finite-range hopping has $C=0$~\cite{Chen2014}. For the bands considered in this paper the Berry curvature is therefore non-uniform whenever it is nonzero. Repeated commutators with the boosts then generate the operators diagonal in $\mathbf k$ with entries $\partial\Omega,\partial^2\Omega,\ldots$, an infinite set of mutually commuting operators that are not central. The boosts thus close into a non-central, infinite-dimensional extension of the Carroll algebra. Its structure is similar to the extensions of the Poincar\'e algebra for a particle in a non-uniform electromagnetic field~\cite{BonanosGomis2010,GomisKleinschmidt2017}, with the Berry curvature playing the role of a magnetic field in momentum space. For the supertranslations, the projected densities $\bar n_{\mathbf q}$ form a deformation of the Girvin-MacDonald-Platzman algebra, which closes only when the Berry curvature and the quantum metric are both independent of $\mathbf k$~\cite{WangSimon2025}. We leave the study of these extended Carroll algebras to future work.

\section{Finite-range hopping: conservation versus closure}
We now give a microscopic perspective on the distinction between flatness and ultra-local Carrollian symmetry established in the main text. For a general finite-range hopping Hamiltonian, we ask when the Hermitian candidates for position-dependent time translations are conserved, and whether their conservation is sufficient to ensure an abelian supertranslation algebra. The projected version of this question leads back to the Berry-curvature and quantum-metric criteria of the main text.

Consider a translation-invariant quadratic Hamiltonian
\begin{equation}
\mathcal{H}=\sum_{\mathbf{x},\mathbf{r}}\psi_{\mathbf{x}+\mathbf{r}}^\dagger T_{\mathbf{r}}\psi_{\mathbf{x}}\,, \quad \text{with} \quad T_{-\mathbf{r}}=T_{\mathbf{r}}^\dagger \,,
 \label{hoppingham}
\end{equation}
on $\mathbb Z^d$, with finitely many orbitals per cell and hopping range $R$. Using $a,b$ to label orbitals, the operator $\psi_{\mathbf{x}a}$ annihilates a particle in orbital $a$ of cell $\mathbf{x}$, and $(T_{\mathbf{r}})_{ab}$ are the elements of the hopping matrix for displacement $\mathbf{r}$. For direct comparison with the physical position densities in the main text, we take the orbitals in a cell to be co-located at $\mathbf{x}$. Writing $\mathcal{H} = \sum_{\mathbf{x},\mathbf{y},a,b} \psi_{\mathbf{x}a}^\dagger H_{\mathbf{x}a,\mathbf{y}b} \psi_{\mathbf{y}b}$, the one-particle Hamiltonian is given by 
\begin{align}
H_{\mathbf{x}a,\mathbf{y}b} = \bra{\mathbf{x},a}H\ket{\mathbf{y},b} = (T_{\mathbf{x} - \mathbf{y}})_{ab}\,.
\end{align}
The corresponding one-particle Bloch Hamiltonian is given by
\begin{align}
H(\mathbf{k}) =\sum_{\mathbf{r}}\eu^{-\iu\mathbf{k}\cdot\mathbf{r}}T_{\mathbf{r}} \,.
\end{align}

\subsection{A microscopic conservation criterion:}
The Heisenberg equation gives $\iu\partial_t\psi_{\mathbf{x}}=\sum_{\mathbf{r}}T_{\mathbf{r}}\psi_{\mathbf{x}-\mathbf{r}}$. Differentiating once more yields
\begin{equation}
 \partial_t^2\psi_{\mathbf{x}}=-\sum_{\mathbf{r}}A_{\mathbf{r}}\psi_{\mathbf{x}-\mathbf{r}}\,, \ \ \text{where} \ \
A_{\mathbf{r}} \equiv \sum_{\mathbf s}T_{\mathbf s}T_{\mathbf{r}-\mathbf s} \,.
 \label{convolution}
\end{equation}
Thus $A_{\mathbf{r}}$ sums the amplitudes of all two-step hopping paths with net displacement $\mathbf{r}$. Their cancellation for every $\mathbf{r}\ne0$ makes the second-order equation local in the cell coordinate, i.e., $\partial_t^2 \psi_{\mathbf{x}} = - A_{\mathbf{0}}\psi_{\mathbf{x}}$. Equivalently,
\begin{align}
 H(\mathbf{k})^2&=A_{\mathbf{0}}
 =\sum_{\mathbf{r}}T_{\mathbf{r}}T_{\mathbf{r}}^\dagger\ge 0 \,,
\label{square}
\end{align}
inferring that all the band energies are flat. We note that this is an additional condition on the hopping matrices. It makes the second-order equation local, but the first-order equation still couples different cells. In particular, $\partial_t\psi (\mathbf{x},0)$ depends on neighboring fields. The condition therefore does not by itself imply independent evolution of each cell or the projected ultra-local symmetry of the main text.

To examine its symmetry content, we associate a candidate generator of position-dependent time translations (supertranslations) with each real cell function $f(\mathbf{x})$. Let us define the operator $F \equiv \sum_{\mathbf{x},a} f(\mathbf{x})\ket{\mathbf{x},a}\bra{\mathbf{x},a}$, or equivalently $F_{\mathbf{x} a,\mathbf{y} b}=f(\mathbf{x})\delta_{\mathbf{x}\mathbf{y}}\delta_{ab}$. Following its Hermitian prescription, take
\begin{equation}
 Q_f=\tfrac12 \left(FH +HF\right),\quad \text{and} \quad
\mathcal{Q}_f=\psi^\dagger Q_f\psi\,.
 \label{charge}
\end{equation}
Here $Q_f$ acts on the one-particle Hilbert space, and $\mathcal Q_f$ is its second-quantized counterpart. Explicitly,
\begin{align}
 \mathcal{Q}_f=\sum_{\mathbf{x},\mathbf{r}}
 \frac{f(\mathbf{x}+\mathbf{r})+f(\mathbf{x})}{2}
 \psi_{\mathbf{x}+\mathbf{r}}^\dagger T_{\mathbf{r}}\psi_{\mathbf{x}}\,.
\end{align}
For $f=1$, the generators reduce to $H$ and $\mathcal{H}$, giving ordinary time translations; $f(\mathbf{x})=x_i$ gives the candidate Carroll boost in direction $i$. For arbitrary $f$, they are candidate supertranslation generators, although conservation and mutual commutation must still be tested. For quadratic operators,
\begin{equation}
 [\mathcal{H},\mathcal{Q}_f]=\psi^\dagger[H,Q_f]\psi\,,
 \qquad [H,Q_f]=\tfrac12[H^2,F].
 \label{conservation}
\end{equation}
Since $[H^2,F]_{\mathbf{x} a,\mathbf{y} b}=(f(\mathbf{y})-f(\mathbf{x}))(H^2)_{\mathbf{x} a,\mathbf{y} b}$, conservation for every $f$ is equivalent to $A_{\mathbf{r} \neq \mathbf{}0}=0$, agreeing with the previous flat-band conditions. 

\subsection{Flat bands and finite caging:}
Under \eqref{square}, an eigenstate satisfies $A_{\mathbf{0}}\ket{u_n(\mathbf{k})}=E_n(\mathbf{k})^2\ket{u_n(\mathbf{k})}$. Hence every energy belongs to the fixed finite set $\{\pm\sqrt{a_n}\}$, where $a_n$ are the eigenvalues of $A_{\mathbf{0}}$, so all bands are flat. Their eigenvectors can nevertheless depend on momentum. For the distinct energies that actually occur, the spectral projectors are
\begin{equation}
 P_E=\prod_{E'\ne E}\frac{H-E'\mathrm{I}}{E-E'}\,,
 \label{polynomial}
\end{equation}
where $\mathrm{I}$ is the identity and the product runs over all distinct energies other than $E$. Finite hopping range then makes $P_E\ket{\mathbf{x},a}$ compactly supported eigenstates spanning the energy-$E$ subspace. Moreover, $\eu^{-\iu tH}=\sum_E \eu^{-\iu tE}P_E$ has range at most $(m-1)R$ for $m$ distinct energies, inferring that an initially localized state remains in a finite region for all time. It can nevertheless move inside that region, and these compact states need not be orthogonal or have disjoint supports. This is why finite caging does not establish the ultra-locality of the main text. Equation \eqref{square} guarantees that all bands are flat, but the converse does not hold. Unlike flatness, this condition depends on the choice of energy zero: adding a constant $c\mathrm{I}$ preserves the flat dispersion but generally makes the squared Bloch Hamiltonian momentum dependent.

\subsection{Projection and the geometric closure condition:}
We now select an isolated nondegenerate flat band of energy $E_0$ and write $P=P_{E_0}$, as in the main text. Independently of \eqref{square}, $HP=E_0P$ gives
\begin{equation}
 \bar{Q}_f\equiv PQ_fP=E_0PFP\,, \quad \text{and} \quad \left[H,\bar{Q}_f\right]=0 \,.
 \label{projectedcharge}
\end{equation}
This is precisely \eqref{qf}: conservation within the band follows from flatness alone. For $E_0\ne0$, the remaining requirement for an abelian supertranslation algebra is
\begin{equation}
 \left[\bar{Q}_f,\bar{Q}_g\right]=E_0^2\left[PFP,PGP\right]=0
 \quad\text{for all }f,g \,.
 \label{projectedclosure}
\end{equation}
Here $G \equiv \sum_{\mathbf{x},a} g(\mathbf{x}) \ket{\mathbf{x},a}\bra{\mathbf{x},a}$ is the multiplication operator for a second cell function $g(\mathbf{x})$. Although $F$ and $G$ commute before projection, their projections need not. Indeed,
\begin{equation}
 [PFP,PGP]=PG(1-P)FP-PF(1-P)GP \,.
\label{projectionleakage}
\end{equation}
For linear $f,g$, this is the projected-position commutator measured by the Berry curvature in \eqref{eq:boostcomm}. For arbitrary $f,g$, Fourier expansion gives the full projected-density condition \eqref{eq:densityalgebra}, equivalent to vanishing generalized minimal metric.

It is important to remark that hopping couples different cells, so shifting the time independently at each cell is generally not a symmetry of the full hopping model. Indeed, the transformation generated by our $\mathcal{Q}_f$ and a literal local time shift act on the original fields as
\begin{equation}
 i[\mathcal Q_f,\psi]= -\tfrac i2(FH+HF)\psi\,,
 \qquad F\dot\psi=-iFH\psi \,.
 \label{fieldaction}
\end{equation}
They agree only when $[H,F]=0$, whereas conservation of $\mathcal Q_f$ requires only $[H^2,F]=0$. Thus, a conserved $\mathcal{Q}_f$ need not generate independent local time shifts of the microscopic fields. The main text instead studies symmetry within a selected flat band. There, $PFHP=PHFP=E_0PFP$, so the two prescriptions agree after projection on both sides, and $\bar{Q}_f=E_0PFP$ is conserved by flatness alone. The abelian supertranslation symmetry additionally requires $[\bar{Q}_f,\bar{Q}_g]=0$ for all $f,g$, as in \eqref{eq:densityalgebra}. 

\subsection{A special case with commuting local charges:}
The one-dimensional, two-orbital nearest-neighbor construction of \cite{Ara:2024:SciPost} is a special case of the setup above. Taking $T_{+1}=T$, $T_{-1}=T^\dagger$, all other hoppings zero, the conditions $A_{r \neq 0} = 0$ enforce the nilpotency, $T^2=0$, and $H(k)^2=TT^\dagger+T^\dagger T$. It can be checked that $[\mathcal{Q}_f, \mathcal{Q}_g] =0$: the additional closure condition \eqref{eq:densityalgebra} is automatic in this special model.

In addition, the bond Hamiltonians $\mathcal{H}_j=\psi_{j+1}^\dagger T\psi_j+\mathrm{h.c.}$ used in \cite{Ara:2024:SciPost} also commute. Consequently the charges $\mathcal{Q}_f^{\mathrm{CLS}} = \sum_jf_j\mathcal H_j$ are conserved and mutually commuting. These charges realize local time shifts on canonical CLS fields combining neighboring cells, while \eqref{fieldaction} instead concerns the original fields $\psi$. In this nearest-neighbor model, $T^2=0$ guarantees both conservation and mutual commutation of the charges. For more general finite-range hopping, however, $H(\mathbf{k})^2=A_{\mathbf{0}}$ guarantees conservation but does not automatically guarantee mutual commutation.

\end{document}